\documentstyle[epsfig,aps]{revtex}

\def\circa#1{\,\raise.3ex\hbox{$#1$\kern-.75em\lower1ex\hbox{$\sim$}}\,}

\newcommand{\ov}{{\cal O}}

\newcommand{\Dsl}{D\hspace{-5.8pt}\vspace{5pt}{ /}}

\newcommand  \f  \varphi
\newcommand \bra {\langle}
\newcommand \ket {\rangle}
\newcommand{\be}{\begin{equation}}
\newcommand{\ee}{\end{equation}}
\newcommand{\ben}{\begin{displaymath}}
\newcommand{\een}{\end{displaymath}}
\newcommand{\ba}{\begin{eqnarray}}
\newcommand{\ea}{\end{eqnarray}}
\newcommand{\ban}{\begin{eqnarray*}}
\newcommand{\ean}{\end{eqnarray*}}
\newcommand{\cro}{\dagger}

\newcommand{\de}{\partial}

\newlength{\www}

\begin{document}
\title{\Large\bf Bloch-Nordsieck Violation in Spontaneously \\
Broken Abelian Theories
\footnote{Work supported in part by EU QCDNET contract
FMRX-CT98-0194 and by MURST (Italy).}}
\author{\large Marcello Ciafaloni}
\address{\it Dipartimento di Fisica, Universit\`a di Firenze and
\\ INFN - Sezione di Firenze and CERN, Geneva}
\author{\large Paolo Ciafaloni}
\address{\it INFN - Sezione di Lecce,
\\Via per Arnesano, I-73100 Lecce, Italy}
\author{\large Denis Comelli}
\address{\it INFN - Sezione di Ferrara,
\\Via Paradiso 12, 44100 Ferrara, Italy}
\maketitle
\vskip0.3cm

\begin{abstract}
We point out that, in a spontaneously 
broken $U(1)$ gauge theory,
inclusive processes, whose  primary particles  are 
mass eigenstates
that do not coincide with the gauge eigenstates,
 are not free of infrared logarithms.
The charge mixing allowed by  symmetry breaking 
and the ensuing Bloch-Nordsieck violation are
 here analyzed in a few relevant cases and 
in particular for processes  initiated by
longitudinal gauge bosons.
Of particular interest is the example of weak
hypercharge   in the 
Standard Model where, in addition, left-right mixing effects arise in
transversely polarized fermion beams.

\end{abstract}

\vskip1.3cm
The planning of TeV scale accelerators has brought attention to the fact
 that the Standard Model, at energies larger than the weak scale, shows
 enhanced double log corrections \cite{3p} of infrared origin, 
even in inclusive observables.
Such enhancements, involving the effective coupling 
$(\alpha_W / 4 \pi) \log^2 (s / M_W^2)$, signal a lack  of compensation
 of virtual corrections with real emission in the $M_W^2 \ll s$ limit, 
due to the non abelian (weak isospin) charges of the accelerator beams.
In other words, the Bloch-Nordsieck (B-N) cancellation theorem \cite{bn}, 
valid in QED, is here violated.

The key point which invalidates the B-N cancellation is the fact that 
gauge boson emission off one incoming beam state changes it into another
 state of the same gauge 
 multiplet (e.g., a neutrino for an incoming electron) and 
the latter happens to have a different cross section off the other beam.
As a consequence, virtual corrections are unable to cancel this 
contribution  except on the average, i.e., by summing over all
 possible beams in the multiplet.

It is usually thought that such phenomenon cannot occur in the abelian case,
 because initial states (the mass eigenstates) are charge eigenstates, which
 do not change during the (neutral) gauge boson emission, so that the
 real-virtual cancellation is valid.

In this note we point out that, in case of spontaneous symmetry breaking,
 the B-N theorem is violated in {\it abelian theories} too.
The point is that, in a broken theory, mass eigenstates can be mixed 
charge states, so that soft boson emission
is off diagonal. For instance, if a  normal Higgs mechanism \cite{higgs}
is assumed,
 longitudinal gauge bosons can occur as (massive) initial states which 
act as mixed charge states and interact with the (similarly mixed) Higgs boson.
As a consequence, longitudinal and Higgs bosons are interchanged during soft
 emission, and the basic noncancellation mechanism is again at work, as in 
the non abelian case illustrated before.
 
In order to understand this point, let us recall the structure of soft
 interactions accompanying a hard process of type 
$\{\alpha_I p_I\}\rightarrow \{\alpha_F p_F\}$, where $I$=1,2, $F$=1,2,...,n, 
and
 $p$'s and $\alpha$'s denote momenta and charge states of initial and final 
asymptotic states, which are mass eigenstates.
The corresponding S-matrix is an operator in the soft
 Hilbert space and  a matrix in the hard labels,
 with form \cite{fk,cm}
\be
 \label{eq:2}
S={\cal U}^F_{\alpha_F\alpha_F'}\! (a_s,a^\cro_s)\;\;\;
S^H_{\alpha_F'\alpha_I'}\! (p_F,p_I)\;\;\;
{\cal U}^I_{\alpha_I'\alpha_I}\! (a_s,a^\cro_s)
\ee
where ${\cal U}^F$ and ${\cal U}^I$ are unitary coherent state operators,
 functionals of the
soft emission operators $a_s,a^\cro_s$.
They take in the abelian case a simple eikonal form \cite{fk} and are
 {\it diagonal} with respect to charge eigenstates, i.e., they have 
a well defined form for each energetic particle of well defined charge.

An inclusive observable is obtained by squaring and summing eq.(\ref{eq:2})
 over soft final states.
In this procedure, the coherent state ${\cal U}^F$ cancels out by unitarity,
 and we are left with the overlap matrix
\be
\label{O}
\ov_{\beta_I \alpha_I}=\;_S
 \bra 0| {\cal U}^{I\cro}_{\beta_I\beta_I'}(S^{H\cro}S^H)_{\beta_I'\alpha_I'}
{\cal U}^I_{\alpha_I'\alpha_I}|0 \ket_S
\ee
where an average over the state with no soft quanta is made in the initial
 state.
We also refer to
 ${\cal O}^H=S^{H\cro}S^H$ as the hard overlap matrix, and we allow in general
$\beta_I\neq \alpha_I$, even if a cross section with initial charge state 
${\alpha_I}$ is diagonal, i.e., 
$ \sigma_{ \alpha_I}  =
\ov_{ \alpha_I \alpha_I}$
(no sum over $\alpha_I$).

The abelian Bloch-Nordsieck cancellation theorem is valid if the initial 
mass eigenstates are also charge eigenstates.
In fact, in such case ${\cal U}^{I}$ is diagonal with respect to the labels 
$\alpha_I=(\alpha_1,\,\alpha_2)$ which represent definite charges,
i.e.,
\be
{\cal U}^I_{\alpha_I'\alpha_I}=\delta_{\alpha_I'\alpha_I}\;
{\cal U}^{\alpha_I p_I},\;\;\;\;\;\;\;
{\cal U}^{\alpha_I p_I}\equiv \Pi_{i=1,2}\;{\cal U}^{\alpha_i p_i}
(a_s,a_s^{\cro})
\ee
Therefore, the inclusive cross section becomes, by eq.(\ref{O})
\be\label{cross}
\sigma_{ \alpha_I} = 
\ov_{ \alpha_I \alpha_I}=
\;_S\bra 0| {\cal U}^{\alpha_I p_I\cro}\,\ov^H_{\alpha_I\alpha_I}\,
{\cal U}^{\alpha_I p_I}|0 \ket_S
\ee
where $\alpha_I$, in both ${\cal U}$ and ${\cal U}^{\cro}$, is now  the
 same set of labels (with no sum).
Since soft operators only occur in the ${\cal U}$'s, the latter commute 
with $\ov^H$, and soft enhancements cancel out by unitarity, in a trivial way.

The above reasoning fails in the non  abelian case, because both ${\cal U}$ and
 $\ov^H$ are
(non commuting) matrices in a non  abelian charge multiplet, and one is 
unable to use the unitarity sum.
But it fails in the abelian case too, if the initial states {\it are not} 
charge eigenstates, as allowed by symmetry breaking.
In such a case, the coherent states are
 not diagonal in the initial labels ${\alpha_I}$, and normally do not commute
 with the hard overlap matrix $\ov^H$.
More precisely, by introducing the mixing matrix ${\cal M}_{A\alpha}$ and 
the overlap matrix $\ov_{AB}$ in the charge eigenstates basis $\{A\}$,
 we obtain:
\be\label{sum}
\sigma_{ \alpha_I} = 
\ov_{ \alpha_I \alpha_I}=\sum_{A,B} {\cal M}_{\alpha_IB}^{\cro}\,\ov_{BA}\,
{\cal M}_{A\alpha_I}
\ee
While soft enhancements cancel out by eq.(\ref{cross}) in the diagonal 
terms $\ov_{AA}$ of the sum (\ref{sum}), they are non vanishing in the off
 diagonal ones $\ov_{AB}$, ($A \neq B$), which are induced by the mixing,
 so that the BN theorem is violated.

We shall illustrate the features above in the example of the longitudinal
 sector of the $U(1)$ Higgs model \cite{higgs}.

\begin{figure}[htb]
\setlength{\unitlength}{1cm}
\begin{picture}(12,7.5)
\put(3.7,0.3){(a)}\put(12.4,0.4){(b)}
\put(0.1,2){\epsfig{file=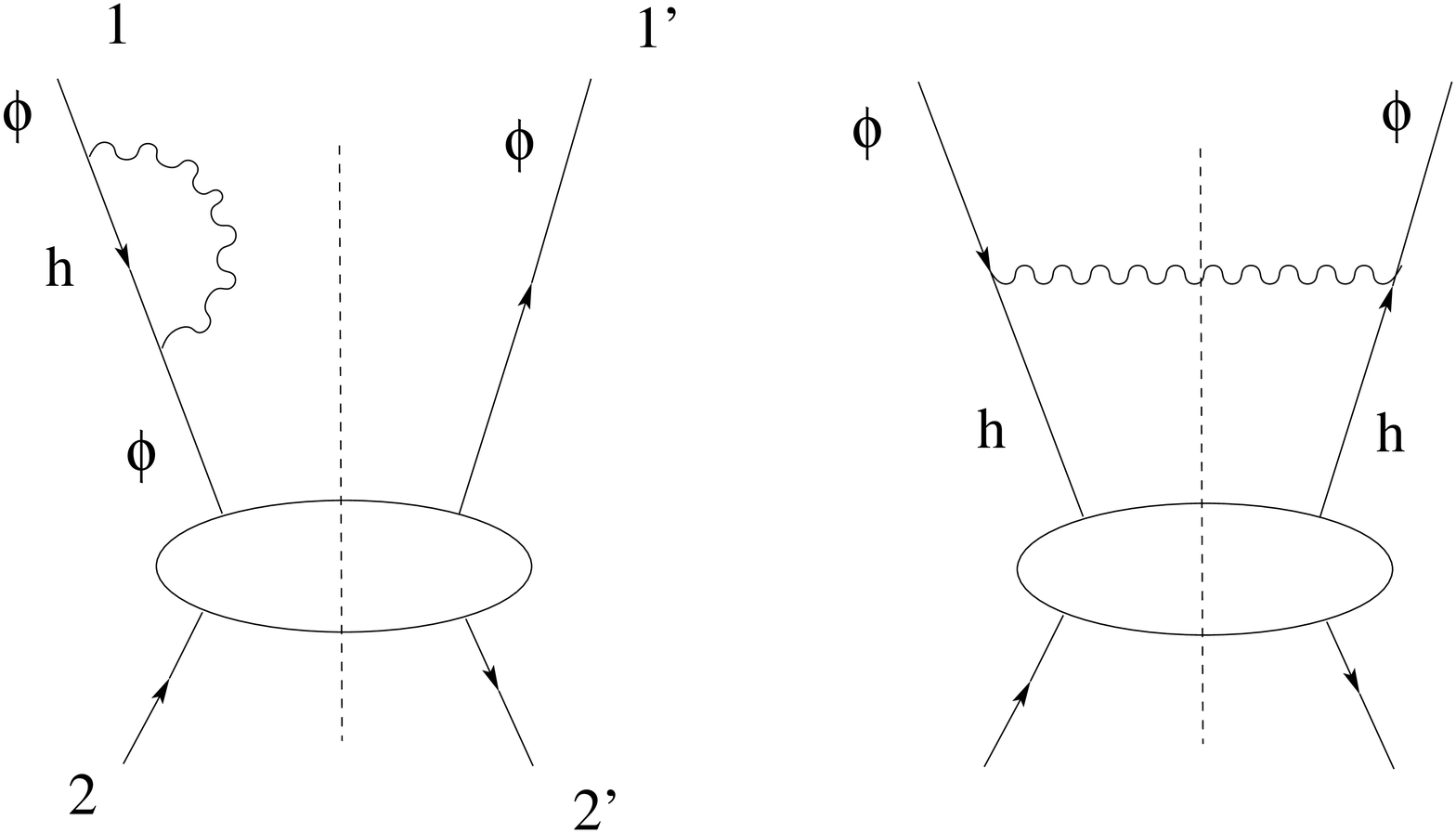,height=4cm}}
\put(10,1){\epsfig{file=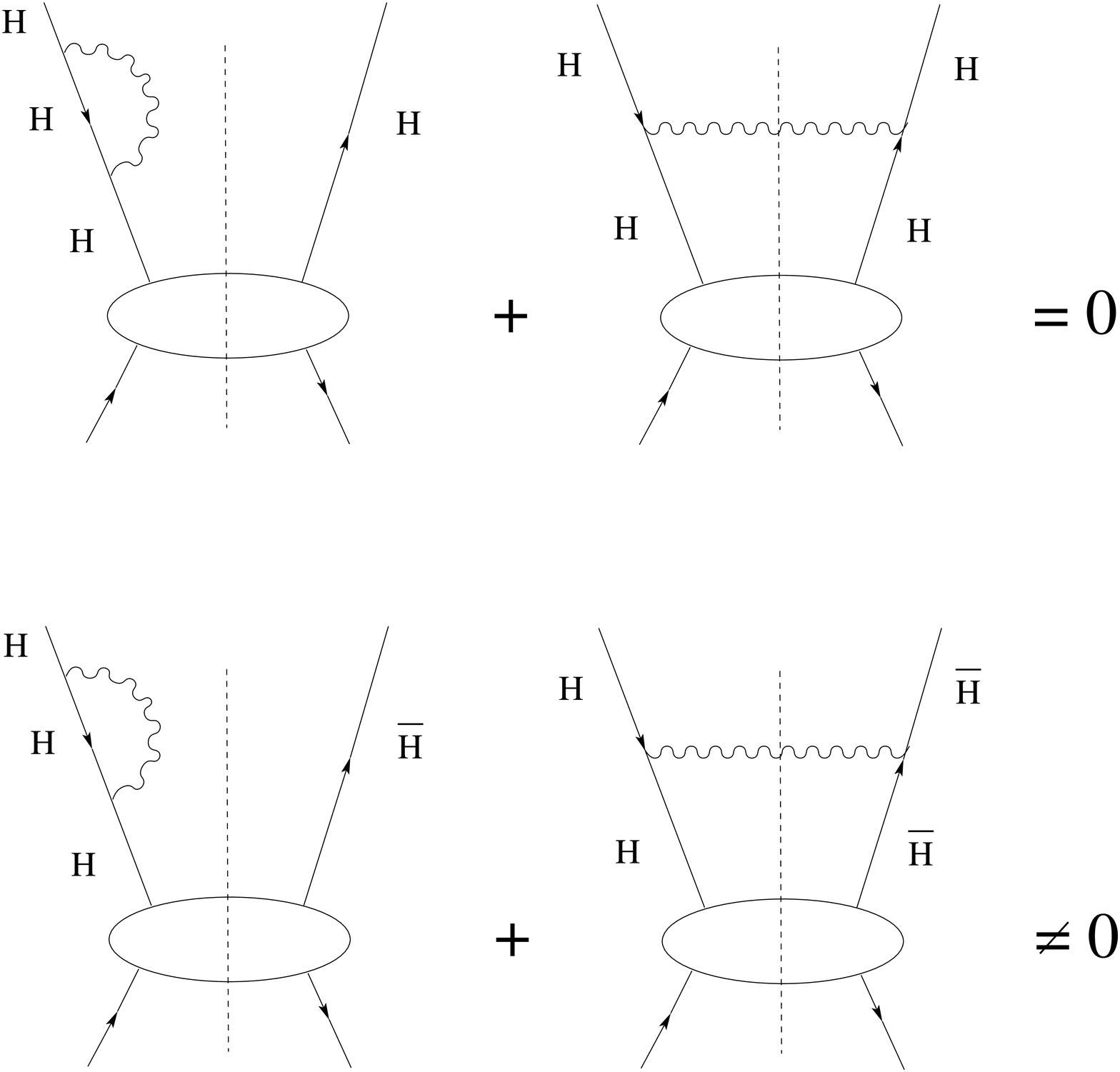,height=6cm}}
\end{picture}
\caption{Picture of radiative corrections to the overlap matrix in (a) the
mass eigenstates basis and (b) the charge eigenstates basis, 
where off-diagonal
matrix elements occur.}
\end{figure}
The Lagrangian in a  't Hooft gauge is
\be
{\cal L}= (D_{\mu}\Phi)^+\,D^{\mu}\Phi
-V[\Phi^+\Phi]-\frac{1}{4} F_{\mu\nu}F^{\mu\nu}-
\frac{1}{2\,\zeta}(\partial_{\mu}A^{\mu}-\zeta\, M\,\Phi)^2
+\bar{\Psi}(i\Dsl-m)\Psi
\ee    
where $\Phi- v /\sqrt{2}=H=(h+i \phi)/ \sqrt{2}$ is the
Higgs field, 
$V$ is the  potential, $M$ is the gauge boson mass, $\zeta$
is the gauge parameter, and charged fermions of mass $m$ have been 
introduced. We also take $M_H \simeq M$ as $h$ field mass, the case 
$M_H \gg M$ being discussed in \cite{abelian}. 

The states we consider are the Higgs boson $h$ and the longitudinal boson 
$A_{\mu}^L \equiv L$ as prepared, e.g., by coupling to initial charges in 
a boson fusion process.
Longitudinal amplitudes are related to the Goldstone boson ones by 
the equivalence theorem \cite{equiv}
\be
\frac{p^{\mu}}{M}\; {\cal M}_{\mu}(p;...)=i{\cal M}(\phi(p);...),
\;\;\;\;\;\;\;\;(p^2\simeq M^2)
\ee
where the remaining amplitude labels are understood.
For this reason, the soft emission properties in the $L/ h$ sector are 
determined by the current 
$h(x) \stackrel{\leftrightarrow}{\de_\mu}\phi(x)$ of the scalar 
sector.

At leading double log level, the emission of a soft gauge boson 
off an energetic longitudinal boson 
$\epsilon_{\mu}^L(p)\sim p_{\mu}/M +O(M/E)$ changes it into a
 Higgs boson, and all subsequent interactions are described by the 
 eikonal current \cite{cm,abelian}
\be\label{JJ}
J^{\mu}_{\alpha\beta}(k)=e\, \frac{p^{\mu}}{p \cdot k}\; { q}_{\alpha\beta}
\;\;\;\;\;\;\;\;(\alpha,\beta=\phi, h)
\ee
where ${ q}=\tau_2$ is just a Pauli matrix connecting the $L/\phi$ 
and $h$ indices.
The peculiarity of eq.(\ref{JJ})
is that it is off diagonal, as expected from the fact that
 mass eigenstates are the mixed charge states $h=1/\sqrt{2}(H+H^{\cro}),\;
\phi=-i/\sqrt{2}(H-H^{\cro})$.

The actual evaluation of double logs in eq. (\ref{cross}) is simplified
 by the remark that the eikonal current
(\ref{JJ}) is conserved in the fixed angle, 
high energy regime $s \gg M^2$ that 
we are investigating.
This means that, by applying the current $J^{\mu}$ to states in the 
overlap matrix
$\bra \beta_1\beta_2|\ov|\alpha_1\alpha_2 \ket$, we have 
\be
k^{\mu}J_{\mu}(k)\,\ov=\sum_i q_i\,\ov=(q_1+q_2-q_{1'}-q_{2'})\ov=0
\ee
where the sum runs over all legs of the overlap matrix, as depicted in 
Fig.1(a).
Furthermore, if we like, we can diagonalize each leg charge $q_i$ by
 reverting to the complex field $H\;(H^{\cro})$ with charge $q=1(q=-1)$.

By using charge conservation for $s\gg M^2$, the total eikonal current
 occurring in the coherent states of eq.(\ref{O})
can be expressed in the simple form
\be
J^{\mu}=e\,\sum_i q_i \frac{p_i^{\mu}}{p_i k}=e\, Q \,(
\frac{p_1^{\mu}}{p_1 k}-\frac{p_2^{\mu}}{p_2 k})
\ee
where $Q=q_1-q_{1'}=q_{2'}-q_{2}$ is the total t-channel charge.
Therefore the eikonal radiation factor, involving the squared eikonal current
$J^{\mu}J_{\mu}$, becomes:
\be \label{QQ}
-Q^2\, \frac{e^2}{8 \pi^3}\int \frac{d^3k}
{2\, \omega_k}\frac{p_1 \cdot p_2}
{p_1 \cdot k\; p_2 \cdot k}\equiv -Q^2\,{\cal L}
\ee
where ${\cal L}=(\alpha / 4 \pi) log^2 (s / M^2)$
is the effective double log coupling mentioned above.
The structure of radiative corrections is then the one depicted in Fig.1(a),
where for each power of $\alpha$ the operator 
$-Q^2=-(q_1-q_1')^2=-2(1-q_1q_1')$ is 
applied. While the virtual corrections are diagonal, the term $q_1q_1'$,
 corresponding
to real emission, exchanges the $h$ and $\phi$ indices on both legs, as 
anticipated before.
If we fix $\alpha_2=\beta_2=L$, and we define $\sigma_{\alpha}=
\sigma_{\alpha L}$, ($\alpha=L,h=\phi,h)$, the action of $q_1q_1'$ on the 
$\alpha$  indices is that of a $\tau_1$ Pauli matrix.
Therefore, by restoring the full radiation factor, we find
\be\label{fin}
\sigma_{\alpha}=\left(e^{-2 {\cal L}(1-\tau_1)}\right)_{\alpha \beta}
\sigma^H_{\beta}
\ee 
where the $\sigma^H$'s are the hard (tree level) cross sections.
The final result (\ref{fin}) is easily recast in the diagonal form
\be\label{LL}
\sigma_{LL}+\sigma_{hL}=\sigma_{LL}^H+\sigma_{hL}^H,\;\;\;\;\;\;\;\;
\sigma_{LL}-\sigma_{hL}=(\sigma_{LL}^H-\sigma_{hL}^H)e^{-4 {\cal L}}
\ee

This means that the average cross section has no radiative corrections,
 while the difference is suppressed by the  form factor corresponding to
 t-channel
 charge $Q^2=4$.
Therefore, at infinite energy, radiative 
corrections equalize the longitudinal
 and Higgs cross sections.

The occurrence of the t-channel charge $Q^2=4$ is related to the basic fact
 that $h$ and $L$ are not charge eigenstates, due to symmetry breaking at low 
energies.
In fact, by rewriting the cross sections in term of the charge eigenstates 
$H$ and $H^{\cro}$, and by using charge conjugation 
invariance  we find
\be\label{ss}\begin{array}{l}
\sigma_{LL}=\sigma_{hh}=\frac{1}{2}(\sigma_{HH}+\sigma_{H\bar{H}})+
Re \ov(H\bar{H}\rightarrow \bar{H}H)\\
\sigma_{Lh}=\frac{1}{2}(\sigma_{HH}+\sigma_{H\bar{H}})-
Re \ov(H\bar{H}\rightarrow\bar{H} H)
\end{array}\ee
where, as in eq. (\ref{sum}) we notice the occurrence of the off diagonal 
overlap matrix elements 
 $\ov$ and $\ov^{\cro}$,  corresponding to the values 
$Q_{tot}=q_1-q_1'=\pm 2$ of the 
total charge in the t-channel (Fig.1(b)).
While the diagonal terms $\sigma_{HH}$ and $\sigma_{H\bar{H}}$ 
 correspond to 
$Q=0$ and have no form factor, the off diagonal ones are suppressed by the
 form 
factor with $Q^2=4$, already found before.
Therefore, from eq.(\ref{ss}) we find the 
expressions
\be\label{la}\begin{array}{l}
\sigma_{LL}(s)=\sigma_{hh}(s)=\frac{1}{2}(\sigma^H_{LL}+\sigma^H_{Lh})+
\frac{1}{2}(\sigma^H_{LL}-\sigma^H_{Lh})\; e^{-4\, {\cal L}}
\\
\sigma_{Lh}(s)=\frac{1}{2}(\sigma^H_{LL}+\sigma^H_{Lh})-
\frac{1}{2}(\sigma^H_{LL}-\sigma^H_{Lh})\; e^{-4\, {\cal L}}
\end{array}\ee
which are equivalent to eq.(\ref{LL}). The derivation based on eq.(\ref{ss})
 makes it clear that this phenomenon is not limited to longitudinal and
 Higgs states, but applies to any mixed charge states which are allowed 
by symmetry breaking.



A final comment is about longitudinal couplings to external charges, e.g.
fermions of mass $m$, 
which make the above effect observable. It is known that, by fermion 
current conservation, longitudinal polarizations are suppressed by a factor 
$M^2/k_T^2$ 
with respect to transverse ones, where $k_T^2$ denotes the boson transverse
momentum, related to its virtuality.
However, if $M\gg m$, then the longitudinal $k_T^2$
distribution is dominated by $k_T^2=O(M^2)$, yielding a cross section 
of the same order as the transverse one \cite{ewa}.
The situation changes in the limit of vanishing symmetry breaking parameter.
 In fact if $M\ll m$, the longitudinal  $k_T^2$
distribution is cut off by $m^2$, rather than $M^2$, thus yielding a cross 
section of relative order $M^2/m^2$ which vanishes, eventually. 
Therefore, in the vanishing $M/m$ limit, gauge symmetry and BN theorem are 
recovered at the same time.

The U(1) Higgs model just discussed is a prototype.
A slightly more complicated example, which is relevant for planned 
accelerators, is electroweak theory itself.
Here the gauge group is $SU(2)_L\otimes U(1)_Y$, and important B-N 
violating corrections are found in the longitudinal sector \cite{abelian} 
of both non abelian and {\it abelian} type.
The latter survive in the formal limit of vanishing isospin coupling
 and have a structure similar to the one illustrated here.

An additional peculiarity of the standard model is that, because of the chiral
nature of the gauge group, massive {\it fermions} of mass $m$ are themselves a 
superposition of left and right states of different weak hypercharge (and
isospin). Therefore, by the general argument of eqs.(\ref{sum}) and (\ref{la}),
{\it abelian} double logs are expected for fermion beams also. If initial
beams are longitudinally polarized, the left/right mixing is small at high
energies, so that the corresponding off-diagonal overlap is suppressed by a 
factor $m^2/s$, and was not explicitly considered before \cite{3p}. 
{\it Transverse} polarizations, however, are a superposition of  left and 
right states with comparable weights: mixing is therefore maximal, as in the 
longitudinal boson case considered so far. The
corresponding off-diagonal overlap provides the azimuthal dependence
\cite{tra} of the inclusive cross section at tree level, and is then affected 
at higher orders by the appropriate double log
form factor (carrying t-channel quantum numbers $Y = t_L = 1/2$ in the
present case). Polarized beam effects provide thus another instance in which
infrared enhancements related to mixing are to be investigated.  


\end{document}